# Lorentz microscopy and small-angle electron diffraction study of magnetic textures in $La_{1-x}Sr_xMnO_3$ (0.15 < $x$ < 0.30): the role of magnetic anisotropy

A. Kotani[1], H. Nakajima[1], K. Harada[1,2], Y. Ishii[1] and S. Mori[1,*]

[1]*Department of Materials Science, Osaka Prefecture University, Sakai, Osaka 599-8531, Japan.*
[2]*Center for Emergent Material Science, the Institute of Physical and Chemical Research (RIKEN), Hatoyama, Saitama 350-0395, Japan.*

Magnetic textures in the ferromagnetic phases of $La_{1-x}Sr_xMnO_3$ for 0.15 < $x$ < 0.30 have been investigated by Lorentz microscopy combined with small-angle electron diffraction experiments. Various types of magnetic textures characterized by stripe, plate-shaped, and cylindrical (magnetic bubble) domains were found. Two distinct types of magnetic stripe domains appeared in the orthorhombic structure with an inversion symmetry of $La_{0.825}Sr_{0.175}MnO_3$, depending significantly on magnetocrystalline anisotropy. Based on *in-situ* observations as functions of temperature and the strength of the external magnetic field, a magnetic field-temperature phase diagram was constructed, showing the stabilization of magnetic bubbles in the ferromagnetic phase of $La_{0.825}Sr_{0.175}MnO_3$.

**Introduction**

A number of intricate magnetic textures such as magnetic bubbles and periodic magnetic stripes, which are induced by magnetic anisotropy and the uniform ferromagnetic order due to the long-range magnetic dipole interaction, have been studied in magnetic materials [1-5]. For example, magnetic bubbles were generated by applying a perpendicular magnetic field to periodic magnetic stripe domains in thin films such as hexagonal barium ferrites and cobalt thin films [1,6,7,8]. In addition, when a magnetic field was applied normal to thin films of some manganese oxides such as $La_{0.875}Sr_{0.125}MnO_3$ and Ru-doped $La_{1.2}Sr_{1.8}Mn_2O_7$, periodic magnetic stripe domains in the ferromagnetic phase turned into magnetic bubbles, which were observed by Lorentz microscopy (LM) [9,10,11].

Manganese oxides with the perovskite structure exhibited a large number of ground states and some anomalous phase transitions such as the magnetic-field-induced insulator-to-metal transition [12]. As shown in Fig. 1(a), $La_{1-x}Sr_xMnO_3$ exhibited a variety of ground states in the electronic and magnetic phase diagram, which depended strongly upon the Sr concentration ($x$) [13, 14]. The phase diagram was mainly divided into two distinct ground states, the ferromagnetic insulating phase with a monoclinic structure (space group (SG): $P2_1/c$) in $x < 0.15$ and the orthorhombic ferromagnetic metallic (FMM) phase (SG: $P$bnm) in $x > 0.15$. Note that the two SGs ($P2_1/c$ and $P$bnm) have an inversion symmetry. Our previous work on LM observation of the magnetic domain structures revealed that the orthorhombic ferromagnetic phase of $La_{1-x}Sr_xMnO_3$ for $x=0.175$ was characterized by nanoscale magnetic stripe domains with antiparallel magnetic moments in adjacent stripe domains [15]. Conversely, in the rhombohedral ferromagnetic phase (SG; $R$-3c), the macroscopic magnetic plate-shaped domains were stabilized. Note that $La_{0.825}Sr_{0.175}MnO_3$ exhibited a rhombohedral-to-orthorhombic transition at approximately 185 K during cooling, accompanying a drastic change in the orientation of the magnetic easy axis from the [111] to the [001] directions.

Recently we found that magnetic domains in $La_{0.825}Sr_{0.175}MnO_3$ depend significantly on the relationship between the magnetic easy axis and the direction of the applied magnetic field. Moreover, $La_{0.825}Sr_{0.175}MnO_3$ with an inversion symmetry showed magnetic cylindrical domains (magnetic bubbles) under the application of the magnetic fields normal to the thin films [16]. Although a large number of magnetic textures such as magnetic skyrmions in magnets *without* inversion symmetry through the relativistic Dzyaloshinski-Moriya (DM) interaction have been studied to date [17-27], few studies of magnetic textures in magnets *with* inversion symmetry have been performed [10]. In this work, we systematically investigate magnetic textures and their changes as functions of temperature and the strength of the external magnetic fields perpendicular to thin films in the ferromagnetic *metallic* phase of $La_{1-x}Sr_xMnO_3$ for $0.15 < x < 0.30$ with inversion symmetry using LM and small-angle electron diffraction (SmAED) experiments.

**Experimental methods**

Single crystals of $La_{1-x}Sr_xMnO_3$ for $x = 0.15, 0.175, 0.20, 0.25$ and $0.30$ were grown by the floating zone method. The crystal structures and orientations in the obtained single crystals were examined by powder X-ray diffraction and back-plate Laue-type X-ray diffraction experiments, respectively. Thin films for LM observation were obtained by grinding with alumina powder and, subsequently, by $Ar^+$-ion sputtering at room temperature. *In-situ* LM experiments were conducted in order to clarify magnetic textures and their changes by applying magnetic fields perpendicular to the thin films. Note that an external magnetic field was applied to the thin films by exciting the objective lens of the conventional TEM (JEM-2100F and JEM-2010). The SmAED experiments were performed at an angular resolution of $10^{-4} \sim 10^{-6}$ rad [28,29,30]. For simplicity, the indices of the crystal planes and electron diffraction spots were represented on the basis of the cubic perovskite structure with $Pm\bar{3}m$. The in-plane magnetizations of the magnetic textures were analyzed with the aid of a phase retrieval technique based on the transport-of-intensity equation (TIE) [31, 32]. Magnetization curves were measured by using vibrating sample magnetometer (VSM) equipped with Quantum Design Physical Property Measurement System (PPMS).

**Results**

Magnetic textures in the ferromagnetic *metallic* phase of $La_{1-x}Sr_xMnO_3$ were investigated by obtaining LM (Fresnel) images and SmAED patterns. Note that LM images are obtained by imaging the intensity distribution at a distance $\Delta f$ below or above the specimen by under- or over-focusing, respectively [33]. Figure 1(b) is an LM image ($\Delta f = -1000$ nm) of $La_{0.8}Sr_{0.2}MnO_3$ at 297 K in the (111) crystal plane of the rhombohedral $R\bar{3}c$ structure, in which the magnetic easy axis is almost parallel to the <111> direction. A magnetic domain structure characterized as a macroscopic magnetic domain with a width of approximately 5 μm appears. As understood in the SmAED pattern of the inset of Fig. 1(b), the primary beam splits into two spots with an angular separation of $\pm\varepsilon$, and there is a diffuse streak between the two split spots. Note that the magnitude of $\varepsilon$ corresponds to approximately 50 μrad. This implies that the macroscopic magnetic domain structure in Fig. 1(b) comprises 180 degree magnetic domains with Bloch-type magnetic domain walls, in which the directions of the magnetic moments of adjacent domains are antiparallel to one another. Note that one-to-one correspondence between the twin structures due to the rhombohedral distortion and the magnetic domain wall can be seen. Conversely, magnetic textures in various crystal planes of the ferromagnetic *metallic* phase with an orthorhombic structure ( *Pbnm* ) of $La_{0.825}Sr_{0.175}MnO_3$ were examined at 100 K. Note that the magnetic easy axis is parallel to the [001] direction in the orthorhombic structure. Figures 1(c) and (d) exhibit LM images with a defocused value of $\Delta f = -300$ nm, exhibiting magnetic textures obtained in the (111) and (001) crystal planes at 100 K, respectively. Magnetic stripe domains comprise regions separated by domain walls that appear as an alternating arrangement of straight lines with bright and dark contrast in the (111) crystal plane, as shown in Fig. 1(c). In the SmAED pattern, there are two split spots indicated by $\pm\varepsilon'$, and there is a diffuse streak between them. From the analysis of the SmAED pattern, magnetic stripe domains are characterized as the 180 degree magnetic domain structure having the Bloch-type magnetic domain walls. Conversely, as shown in Fig. 1(d), magnetic stripe domains were found in the (001) crystal planes, which are perpendicular to the magnetic easy axis along the [001] direction. In the LM image, alternative bright and dark lines can be seen. Unlike the SmAED patterns of Figs. 1(b) and (c), the SmAED pattern in Fig. 1(d) shows the presence of the primary spot in the reciprocal space, as indicated by arrow O, in addition to two split spots ($+\alpha$ and $-\alpha$) due to magnetic deflection by the magnetic moment in the sample. The presence of the primary spot implies that there is a magnetic component perpendicular to the (001) crystal plane. Based on the analysis of both SmAED patterns and LM images, the spatial distributions of magnetization in each domain are illustrated schematically in Figs. 1(e)-(g). The significant point of difference between the three distinct types of magnetic domain structures in Figs. 1(b)-(d) is that there exists magnetic moment components perpendicular to thin films only in the magnetic domain structure of the (001) crystal plane of $La_{0.825}Sr_{0.175}MnO_3$.

Changes in the magnetic domains due to the application of an external magnetic field were examined by *in-situ* LM observation. Figure 2 shows the magnetic field-dependence of a typical 180 degree ferromagnetic domain structure in the (111) crystal plane of the rhombohedral structure in $La_{0.8}Sr_{0.2}MnO_3$. When an external magnetic field is applied perpendicular to the (111) plane, magnetization in the magnetic domains is gradually saturated by the Zeeman effect, and magnetic contrasts almost disappear in the field at 80 mT. In contrast, it is found that magnetic stripe domains in the (111) crystal plane of the orthorhombic structure of $La_{0.825}Sr_{0.175}MnO_3$ changed in a different manner from those of $La_{0.8}Sr_{0.2}MnO_3$. Changes of magnetic stripe domains in the orthorhombic structure of $La_{0.825}Sr_{0.175}MnO_3$ as functions of the strengths of external magnetic fields perpendicular to the thin films are shown in Fig. 3. At $B$ =34 mT, the magnetic stripe domains were pinched off and elliptical isolated magnetic domains appeared at their edges, as shown in the square region of Fig. 3(b). Note that the magnified LM image and the schematic illustration of the elliptical isolated magnetic domain are displayed in Fig. 3(d), and it is found that this domain is similar to the ones observed in the ferromagnetic phase of $La_{0.875}Sr_{0.125}MnO_3$ [9]. As the strength of the magnetic fields increased, the magnetic stripe and elliptical isolated magnetic domains finally disappeared at $B$ =100 mT in Fig. 3(c). As the strength of the magnetic field decreased from $B$ = 100 mT, the magnetic stripe domains appeared reversibly around $B$ = 50 mT with thermal hysteresis.

As shown in Figs. 1(c) and (d), two distinct types of magnetic stripe domains appeared in the (111) and (001) planes of the orthorhombic structure of $La_{0.825}Sr_{0.175}MnO_3$. The first were the 180 degree magnetic domains with in-plane magnetizations antiparallel to those of adjacent domains, and the second were the magnetic stripe domains with magnetizations perpendicular to the (001) crystal planes. Thus, we applied magnetic fields perpendicular to two distinct types of magnetic stripe domains and investigated their changes as functions of the strength of the external magnetic fields. Unlike the changes observed in the magnetic stripe domains in the (111) plane under the application of an external magnetic field, as shown in Fig. 3, we found the appearances of type-I and -II magnetic bubbles clockwise (CW) and counterclockwise (CCW) spin rotations by applying magnetic fields perpendicular to the magnetic stripe domains in the (001) crystal planes [16]. Note that only the stripe domains in the (001) plane changed into the bubble-type magnetic domains by applying external magenta field perpendicular to the thin film. The transition process from the magnetic stripe domains to type-I and -II magnetic bubbles have been reported in our recent work [16]. Note that the type-I and -II magnetic bubbles are in accordance with the definition in Ref. 3, and they are illustrated in Fig. 4(a). In order to confirm the spin configurations of type-I and -II magnetic bubbles found in the ferromagnetic phase of $La_{0.825}Sr_{0.175}MnO_3$, we analyzed the LM images showing type-I and -II magnetic bubbles with the aid of the so-called phase retrieval TIE method. Figure 4 represents the LM images taken at 80 K in the orthorhombic structure of $La_{0.825}Sr_{0.175}MnO_3$ under a perpendicular magnetic field of approximately 570 mT. The images show three kinds of magnetic domain patterns. Under a zero field, the magnetic stripe domains were observed. As the magnetic field was applied perpendicular to the thin film, the transition to the magnetic bubbles took place in a weak magnetic field of 400 mT. Three kinds of magnetic bubbles appeared at approximately 570 mT, as shown in Fig. 4(b). Note that the LM image in Fig. 4(b) was obtained at a defocused value of $\Delta f = -0.2$ mm. The spatial distribution of type-I magnetic bubbles with dark and bright contrast at the center stemmed from the CW and CCW spin rotations of their domain walls. This randomness of the chirality of the magnetic bubbles contrasts with the single-chirality in DM helimagnets [18,19]. Type-II magnetic bubbles also exhibit random chirality with CW and CCW spin rotations. The LM image in Fig. 4(c) is obtained at the defocused value of $\Delta f = +0.2$ mm. The dark and bright contrast of the magnetic bubbles was completely reversed, as understood by comparing Figs. 4(b) and 4(c). We analyzed the LM image showing the magnetic bubbles (Figs. 4(b) and 4(c)) using the TIE analysis method, which reproduced the orientation and magnitude of the in-plane magnetization [32-36]. Note that the color mapping in the inset of Fig. 4(e) exhibited the direction of the in-plane magnetization. The directions of the in-plane magnetizations of type-I and -II magnetic bubbles are visualized in Fig. 4(d). The black color at the centers of type-I magnetic bubbles shows that no in-plane magnetization was present, thus magnetization is assumed to be vertical to the thin film with magnetization being parallel to the [001] direction. The magnified images of type-I magnetic bubbles with left- and right-handed spin rotations and those of type-II magnetic bubbles are shown in Figs. 4(e), (f) and (g), respectively, together with color representations of the in-plane magnetizations calculated using the TIE analysis method.

To clarify the stabilizing region of the magnetic bubbles, we investigated the disappearance of magnetic bubbles as a function of the temperature and the strength of the external magnetic fields. Figure 5 shows a series of changes in the magnetic bubbles on warming from 90 K at a constant magnetic field of 500 mT. The average diameter of the magnetic bubbles was approximately 300 nm at 90K. Upon increasing the temperature from 90 K, magnetic bubbles shrunk gradually from 300 to 200 nm at 165 K and disappeared above 185 K, as shown in Fig. 5(d). Note that a structural phase transition from rhombohedral to orthorhombic structure occurred at approximately 185 K, during which the magnetic easy axis changed from the [001] direction to the [111] direction. Conversely, when the temperature decreased below 200 K, magnetic bubbles appeared reversibly around 185 K and the diameters of the magnetic bubbles increased to 300 nm at 90 K. These reversible changes with respect to temperature at a constant strength of $B=$ 500 mT indicate that magnetic bubbles existed as a stable state in the definite range of the *B-T* phase diagram.

The stability of the magnetic bubbles as a function of the strength of external magnetic fields was investigated at constant temperatures of 90, 120 and 165 K in the orthorhombic phase. The average diameter (*r*) of the magnetic bubbles was approximately 300 nm at $T = 90$ K and $B = 400$ mT. When the strengths of the external magnetic fields increased, the diameters (*r*) of the magnetic bubbles decreased and until the magnetic bubbles disappeared at $B = 720$ mT. Similar tendencies were exhibited at $T = 120$ and 165 K, as shown in Fig. 6(a). Based on these experimental results, the *B-T* phase diagram that shows the stability of the magnetic bubbles is presented in Fig. 6(b). Note that the specimen thickness in the LM observation is approximately 100 ~ 150 nm. The range in which magnetic bubbles exist stably in the *B-T* phase diagram depends strongly upon the specimen thickness, as in the *B-T* phase diagram of the magnetic skyrmion in some helical magnets [18].

**Discussion**

Here let us briefly discuss the magnetic bubbles found in the ferromagnetic *metallic* phase of $La_{1-x}Sr_xMnO_3$. Formation of magnetic bubbles depends strongly upon magnetic anisotropy. The magnetic easy axis is parallel to the [001] direction in the orthorhombic phase and to the <111> direction in the rhombohedral phase. The magnetic 180 degree domain structure with in-plane magnetization is formed in the (111) crystal plane of the rhombohedral phase, as shown in Fig. 2, although the magnetic easy axis is parallel to the incidence direction of the electron beam. This clearly implies that the magnetocrystalline anisotropy in the rhombohedral structure is relatively low and that the shape anisotropy is more dominant. As a result, magnetic bubbles are not formed by the application of external magnetic fields perpendicular to thin films in the rhombohedral structure. Conversely, Nagai *et al*. obtained a magnetic anisotropy constant of $K_u = 2.1 \times 10^5$ J/m$^3$ in the ferromagnetic phase of La$_{0.875}$Sr$_{0.125}$MnO$_3$ at 100 K [9], which is larger than $K_u \sim 1.0 \times 10^4$ J/m$^3$ for the ferromagnetic phase of La$_{0.7}$Sr$_{0.3}$MnO$_3$ at 300 K. Moreover, it is believed that this difference arises from the fact that magnetic anisotropy in the ferromagnetic phase is enhanced by orbital ordering.

We examined variations of magnetic anisotropy constants ($K_u$) as a function of temperature in the ferromagnetic phase with the orthorhombic structure of La$_{0.825}$Sr$_{0.175}$MnO$_3$. Figure 7 shows changes of magnetization curves as a function of temperature in the ferromagnetic state of La$_{0.825}$Sr$_{0.175}$MnO$_3$ with the orthorhombic structure. Magnetic fields were applied along the *c* axis and the *ab* plane perpendicular to the *c* axis. Values of saturated magnetization at *T*= 90 K, 120 K and 165 K are approximately 3.7, 3.6 and 3.4 $\mu_B$, respectively. From magnetization curves at *T*= 90 K, 120 K and 165 K, we obtained the anisotropic magnetic field ($H_k$) in the ferromagnetic phase with the orthorhombic structure. Note that H$_k$ is defined as the critical value of the magnetic field, above which the difference in magnetization between H //*c* and H//*ab* is less than 2 %. Values of H$_k$ at *T*= 90 K, 120 K and 165 K are 5.3, 5.2 and 4.8 kOe, respectively. The $K_u$ values obtained based on the anisotropic magnetic field *T*= 90 K, 120 K and 165 K are, respectively, $1.52 \times 10^5$ J/m$^3$, $1.47 \times 10^5$ and $1.30 \times 10^5$ J/m$^3$, which are larger than that of rhombohedral structures such as La$_{0.7}$Sr$_{0.3}$MnO$_3$. Based on these considerations, it is concluded that the magnetic stripe domains in the orthorhombic structure with magnetic moments perpendicular to the thin film are transformed into magnetic bubbles by the application of external magnetic fields perpendicular to the thin film.

## Conclusion

We revealed magnetic textures in the ferromagnetic phases with the inversion symmetry of La$_{1-x}$Sr$_x$MnO$_3$ for $0.15 < x < 0.30$. These textures were characterized as stripes, plate-shapes and magnetic bubbles. By applying external magnetic fields perpendicular to the (001) crystal plane of the ferromagnetic *metallic* phase in La$_{0.825}$Sr$_{0.175}$MnO$_3$, the magnetic stripe domains changed into type-I and -II magnetic bubbles. Magnetic states with type-I and -II magnetic bubbles in the ferromagnetic *metallic* phase exist stably in the temperature-magnetic-field phase diagram.

**ACKNOWLEDGEMENTS**

This work was partially supported by a Grant-in-Aid for Scientific Research from the Ministry of Education, Culture, Sports, Science and Technology of Japan. The authors would like to thank Dr. Y. Togawa and Dr. T. Koyama for useful discussion.

Figure captions

Figure 1. (color online)

(a) Electronic phase diagram of $La_{1-x}Sr_xMnO_3$, in which O, R and M denote the orthorhombic, rhombohedral and monoclinic crystal systems and PM, CA and FM denote the paramagnetic, canted antiferromagnetic, and ferromagnetic phases, respectively [13]. (b) Ferromagnetic domain structures in the (111) crystal plane of the rhombohedral structure of $La_{0.8}Sr_{0.2}MnO_3$ at room temperature. (c), (d) Fresnel images showing magnetic stripe domains in the (c) (111) and (d) the (001) planes of the orthorhombic structure of $La_{0.825}Sr_{0.175}MnO_3$ at 100 K. In the insets of (b)-(d), SmAED patterns are shown. (e)-(f) Schematic depictions of the magnetic domains with the in-plane antiparallel magnetization in the (111) plane of (e) the rhombohedral and (f) the orthorhombic structures, respectively. (g) Schematic description of the magnetic domains with magnetic moments normal to the (001) plane in the orthorhombic sturcture. Red arrows represent the directions of magnetization in each magnetic domain. The defocused values in (b), (c) and (d) are $\Delta f$ = -1000 nm, -300 nm, -300 nm, respectively.

Figure 2. (color online)

Changes of magnetic stripe domains in the (111) crystal plane as a function of the strength of magnetic field perpendicular to the thin film of $La_{0.8}Sr_{0.2}MnO_3$. The strengths of the magnetic field were (a) 20 mT, (b) 50 mT and (c) 80 mT. Red arrows represent the directions of the magnetic moments in each magnetic domain.

Figure 3. (color online)

Changes of magnetic stripe domains in the (111) crystal plane as a function of the strength of the magnetic field perpendicular to the thin film of $La_{0.825}Sr_{0.175}MnO_3$. The strengths of the magnetic field were (a) 20 mT, (b) 34 mT and (c) 100 mT. (d) Magnified ellipsoidal magnetic domain obtained from the square region of (c) and schematic illustration of the ellipsoidal magnetic domain. The red arrow shows the direction of the magnetic moment inside the ellipsoidal magnetic bubble and the magnetic moment is perpendicular to thin film outside.

Figure 4. (color online)

(a) Schematic illustrations of type-I and -II magnetic bubbles. Red arrows represent the direction of the magnetic moment. (b), (c) Magnetic bubbles generated by applying external magnetic fields perpendicularly to a thin film of $La_{0.825}Sr_{0.175}MnO_3$ in the (001) crystal plane. The strength of the applied magnetic field is approximately 570 mT. The images were taken at the defocused values of (b) $\Delta f$= -0.2 mm and (c) $\Delta f$=+0.2 mm. In (c), the type-I magnetic bubbles with left-handed and right-handed spin rotations, and the type-II magnetic bubbles are indicated by capital letters A, B, and C, respectively. (d)-(g) Color representations of the in-plane magnetization calculated with the aid of the TIE analysis method. The directions of the magnetization are (e) type-I with left-handed spin rotation, (f) type-I with right-handed spin rotation, and (g) type-II magnetic bubbles, which were indicated by color mapping and vectorial representations by white arrows.

Figure 5. (color online)
Temperature variations of the magnetic bubble domains in the (001) crystal plane under a constant magnetic field of 500 mT normal to the thin film of $La_{0.825}Sr_{0.175}MnO_3$. The images were obtained at (a) 90 K, (b) 140 K (c) 165 K and (c) 200 K, respectively.

Figure 6 (color online)
(a) Variations of the average diameters ($r$) of magnetic bubbles as functions of strength of the applied magnetic fields at 90, 120, and 165 K, respectively. (b) The $B$-$T$ phase diagram constructed on the basis of the experimental results in this study, in which circles indicate experimentally-measured points.

Figure 7 (color online)
Magnetization curves of $La_{0.825}Sr_{0.175}MnO_3$ at 90 K, 120 K and 165 K. Magnetic field ($H$) was applied parallel (red circles) or perpendicular (blue circles) to the $c$ axis. Each solid triangle shows the anisotropic field.

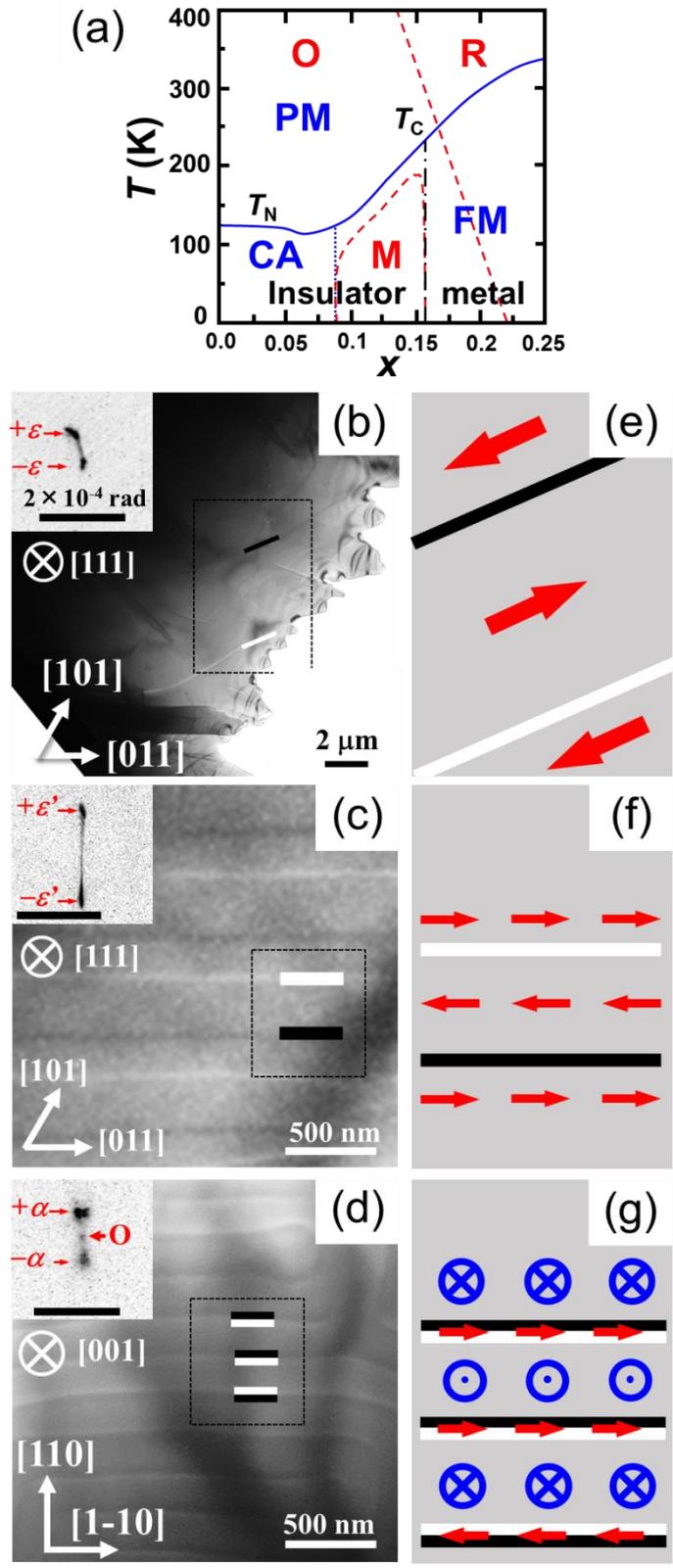

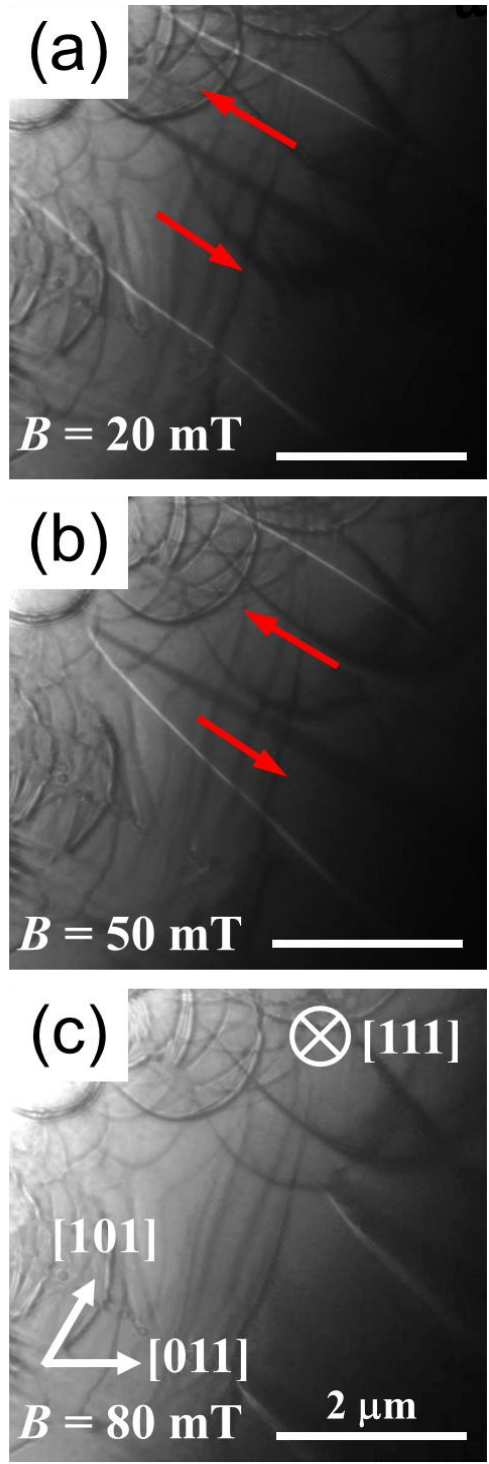

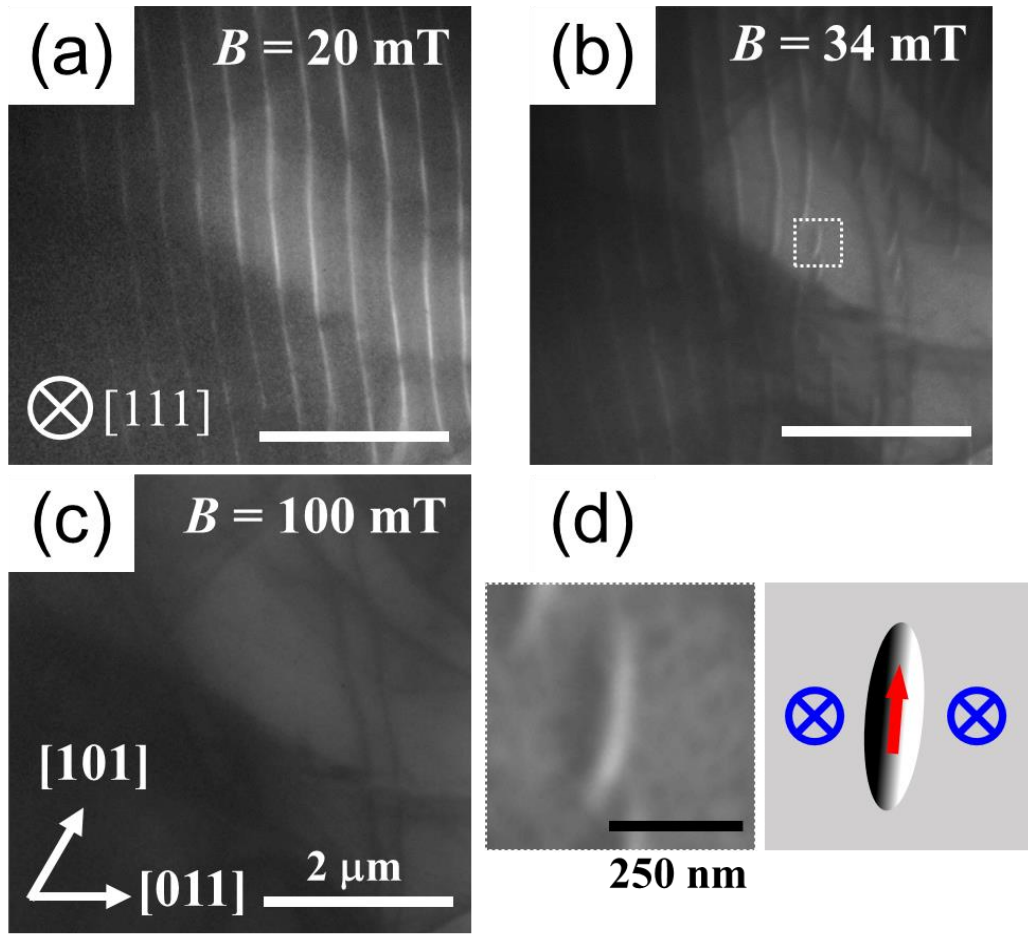

Figure 3, A. Kotani *et al.*

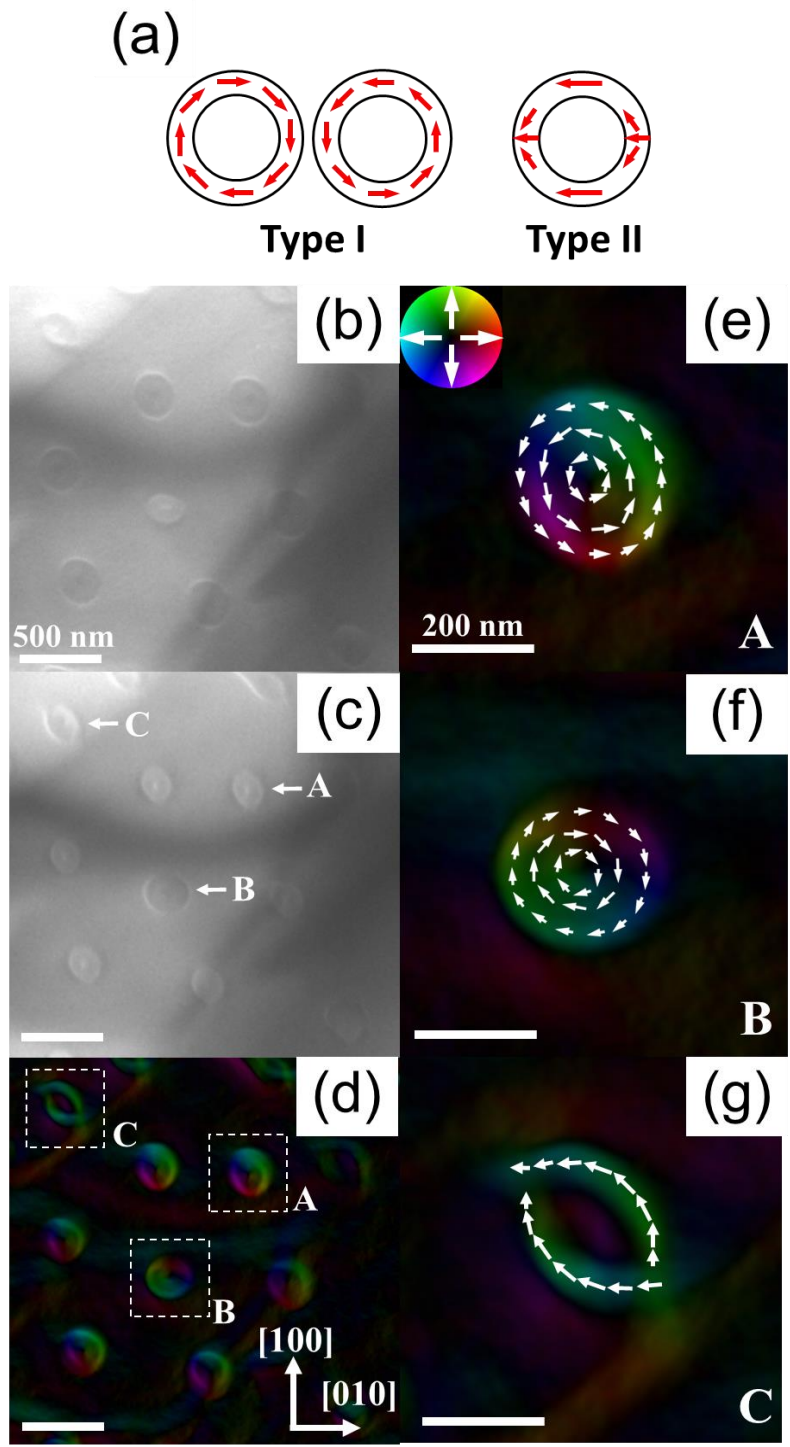

Physical Review B

Figure 4, A. Kotani *et al.*

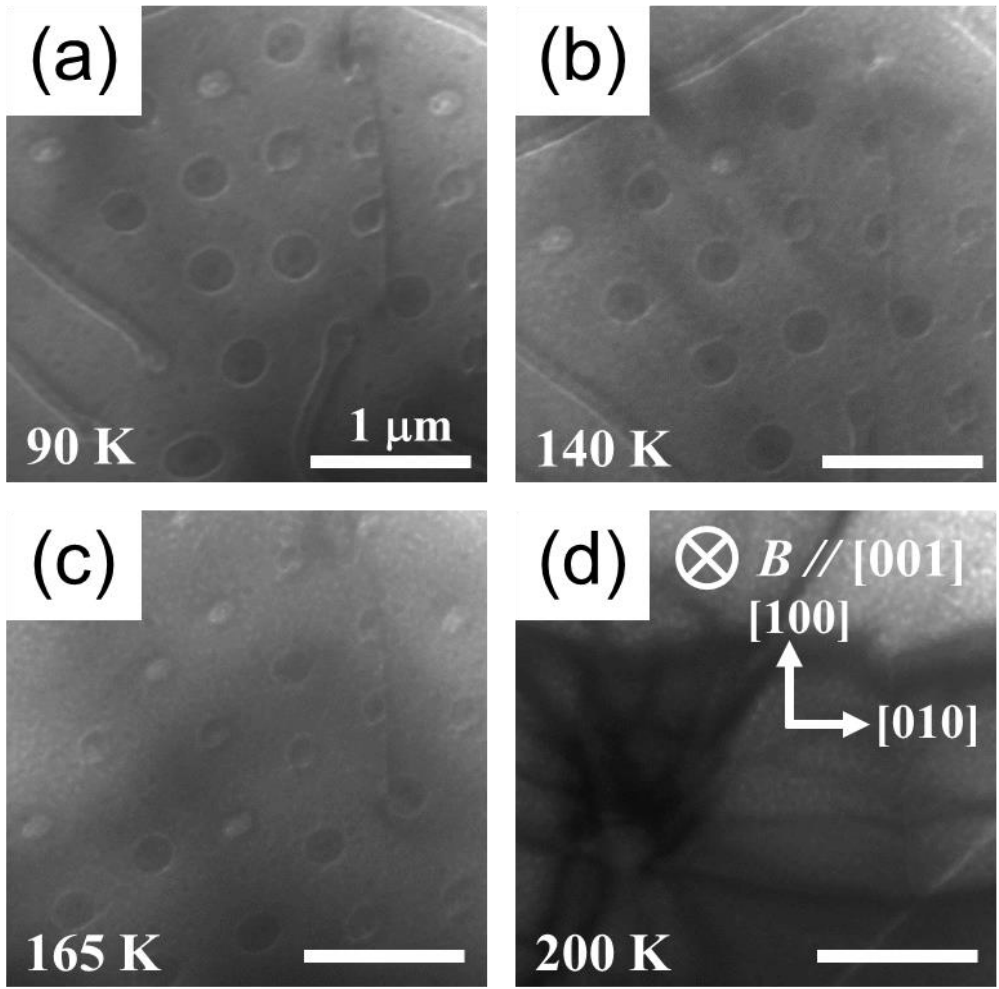

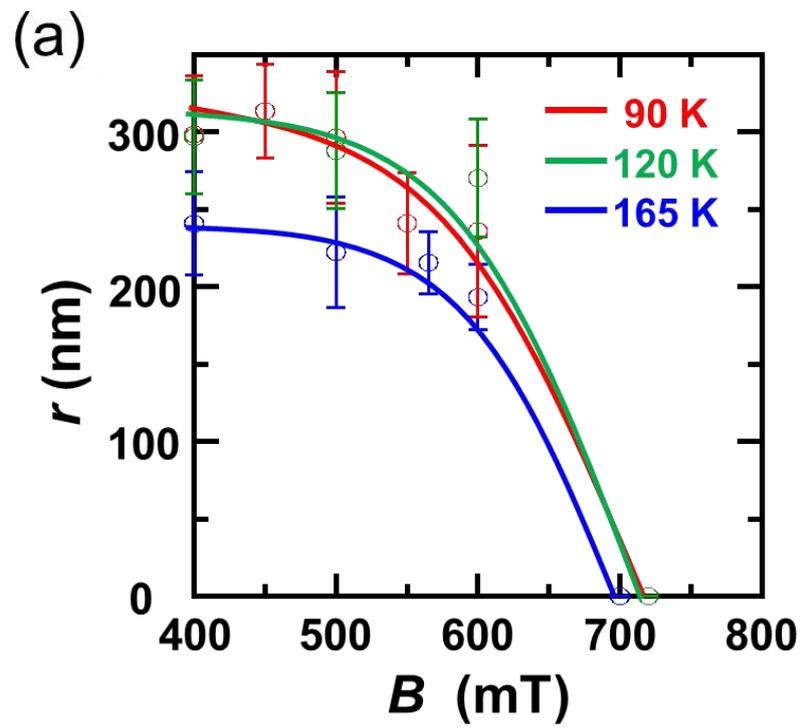
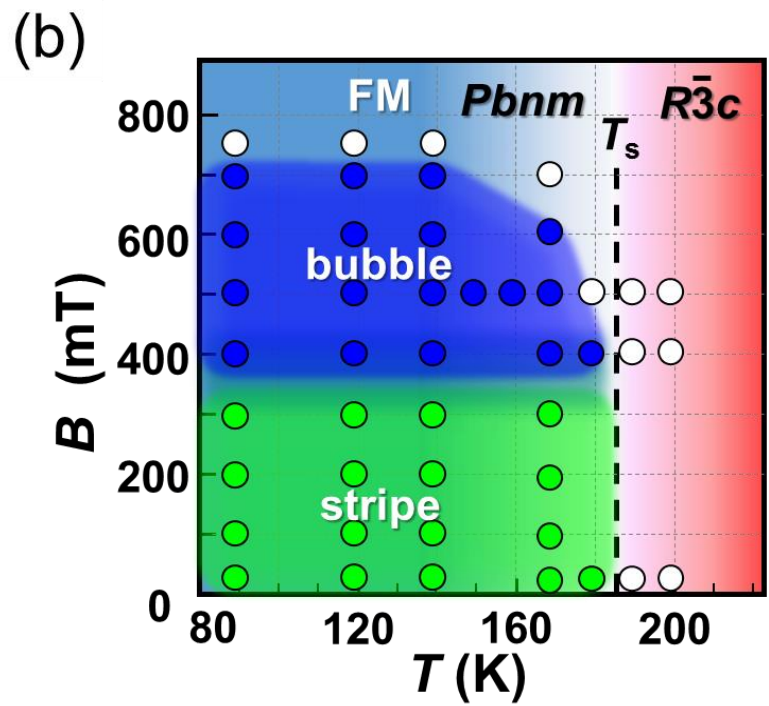

Figure 6, A. Kotani *et al*.

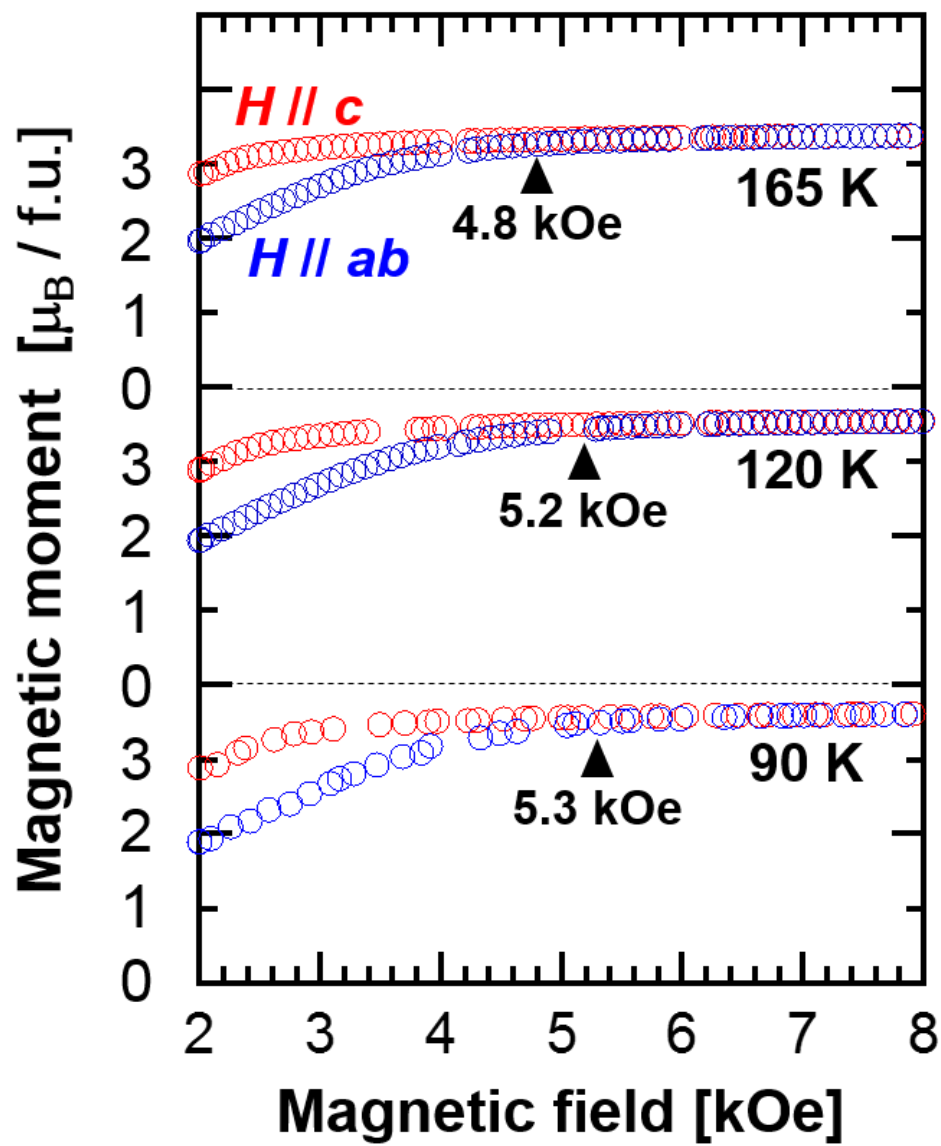

Figure 7, A. Kotani *et al.*